\documentclass{article}
\usepackage[final,nonatbib]{neurips_2018}
\usepackage{cite}
\usepackage{url}
\usepackage[colorinlistoftodos]{todonotes}
\usepackage{times}
\usepackage{gensymb}
\usepackage{lineno}
\usepackage{color}

\title{Contactless Cardiac Arrest Detection \protect \\ Using Smart Devices}

\author
{Justin Chan,$^{1}$ Thomas Rea$^{2,3}$, Shyamnath Gollakota,$^{1\ast}$ Jacob E. Sunshine$^{4\ast}$\\
\\
\normalsize{$^{1}$Paul G. Allen School of Computer Science and Engineering, University of Washington, WA}\\
\normalsize{$^{2}$Division of General Internal Medicine, University of Washington, WA}\\
\normalsize{$^{3}$Medic One, Emergency Medical Services, King County, Seattle, WA}\\
\normalsize{$^{4}$Department of Anesthesiology \& Pain Medicine, University of Washington, WA}\\
\\
\normalsize{$^\ast$To whom correspondence should be addressed:  jesun@uw.edu, gshyam@uw.edu}
}

\begin{document}



\maketitle

\begin{abstract}
Out-of-hospital cardiac arrest (OHCA) is a leading cause of death worldwide. Rapid diagnosis and initiation of cardiopulmonary resuscitation (CPR) is the cornerstone of therapy for victims of cardiac arrest. Yet a significant fraction of cardiac arrest victims have no chance of survival because they experience an unwitnessed event, often in the privacy of their own homes. An under-appreciated  diagnostic element  of  cardiac arrest  is  the presence of agonal breathing, an audible biomarker and brainstem reflex that arises in the setting of severe hypoxia. Here, we demonstrate that a support vector machine (SVM) can classify agonal breathing instances in real-time within a bedroom environment. Using real-world labeled 9-1-1 audio of cardiac arrests, we train the SVM to accurately classify agonal breathing instances. We obtain an area under the curve (AUC) of 0.998 $\pm$ 0.004 and an operating point with an overall sensitivity and specificity of 97.03\% (95\% CI: 96.62 -- 97.41\%) and 98.20\% (95\% CI: 97.87 -- 98.49\%). We achieve a false positive rate between 0\% -- 0.10\% over 82 hours (117,895 audio segments) of polysomnographic sleep lab data that includes snoring, hypopnea, central and obstructive sleep apnea events. We demonstrate the effectiveness of our contactless system in identifying real-world cardiac arrest-associated agonal breathing instances and successfully evaluate our classifier using commodity smart devices (Amazon Echo and Apple iPhone).
\end{abstract}
\section*{Introduction}

Out-of-hospital cardiac arrest (OHCA) is a leading cause of death worldwide\cite{myat2018out} and in North America accounts for nearly 300,000 deaths annually.\cite{mcnally2011out} A relatively under-appreciated diagnostic element of cardiac arrest is the presence of a distinctive type of disordered breathing: agonal breathing.\cite{graham2015strategies,rea2005agonal} Agonal breathing, which arises from a brainstem reflex in the setting of severe hypoxia\cite{poets1999gasping,lumsden1923observations}, appears to be evident in approximately half of cardiac arrest cases reported to 9-1-1. Agonal breathing indicates a relatively short duration from arrest and has been associated with higher survival rates.\cite{hauff2003factors,clark1992incidence,baang2003interaction} 
Sometimes reported as ``gasping" breaths, agonal respirations may hold potential as an audible diagnostic biomarker, particularly in unwitnessed cardiac arrests that occur in a private residence, the location of 2/3 of all OHCAs.\cite{van2017multistate,eisenberg1986identification}

The widespread adoption of smartphones and smart speakers (projected to be in 75\% of US households by 2020\cite{gartner}) presents a unique opportunity to identify this audible biomarker and connect unwitnessed cardiac arrest victims to Emergency Medical Services (EMS) or others who can administer cardiopulmonary resuscitation (CPR). In this study, we hypothesized that existing commodity devices (e.g., smartphones and smart speakers) could be used to accurately identify OHCA-associated agonal breathing instances in a domestic setting. As initial proof-of-concept, we focus on a relatively controlled environment, the bedroom, which is the location of the majority of OHCA events that occur within a private residence.\cite{KIYOHARA2019,eisenberg1986identification} A key challenge to algorithm development for this purpose is accessing real-world data on agonal breathing; agonal breathing events are relatively uncommon, lack gold-standard measurements and cannot be reproduced in a lab because of their emergent nature. To overcome this challenge, we leverage a unique data source, 9-1-1 audio of confirmed cardiac arrest cases, which can include agonal breathing instances captured during the call. As our negative dataset, we use ambient household noise and audio from polysomnographic sleep studies, which include data that share similar characteristics to agonal breathing such as snoring and apnea. Using real-world audio of agonal instances from OHCA events, we evaluate (1) whether a support vector machine (SVM) can be trained to  detect OHCA-associated agonal breathing instances in a bedroom and sleep setting and (2) whether the SVM can be deployed and accurately classify agonal breathing audio in real-time using  existing commodity smartphones and smart speakers.

\begin{figure}
    \centering
    \includegraphics[width=\textwidth]{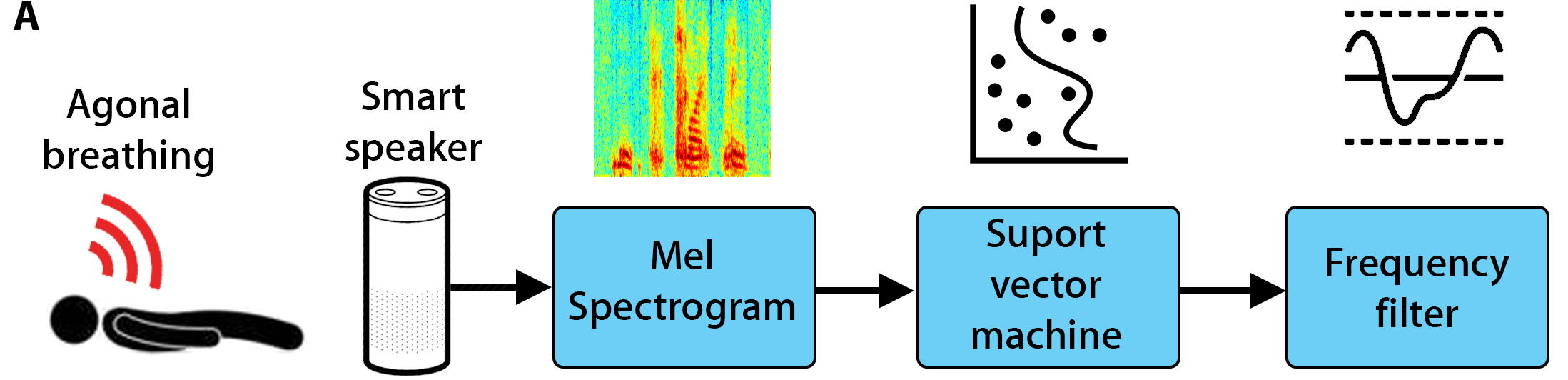}
    \includegraphics[width=.24\textwidth]{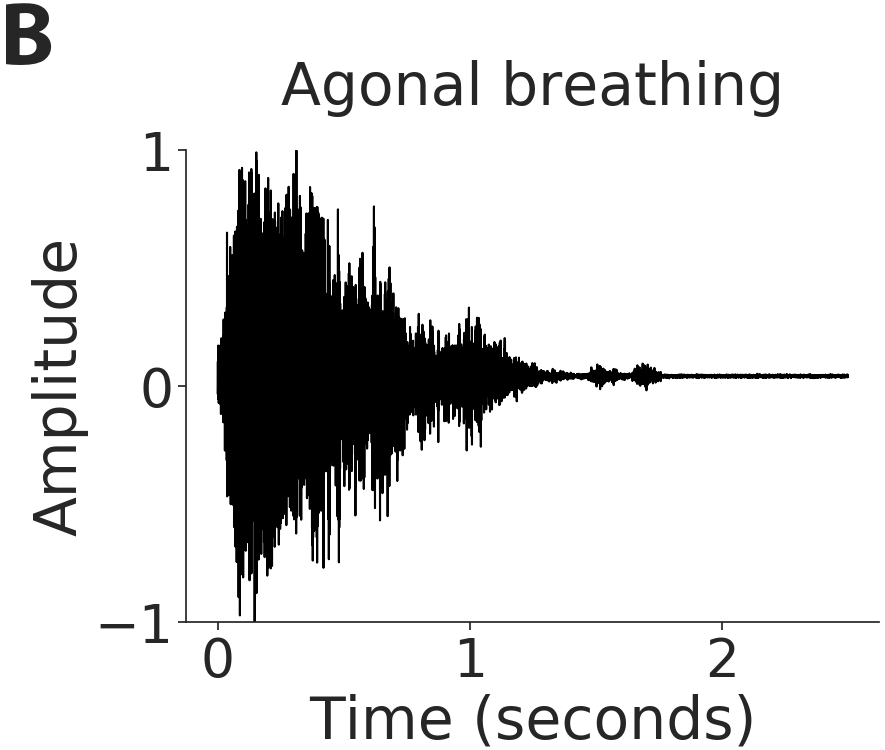}
    \includegraphics[width=.24\textwidth]{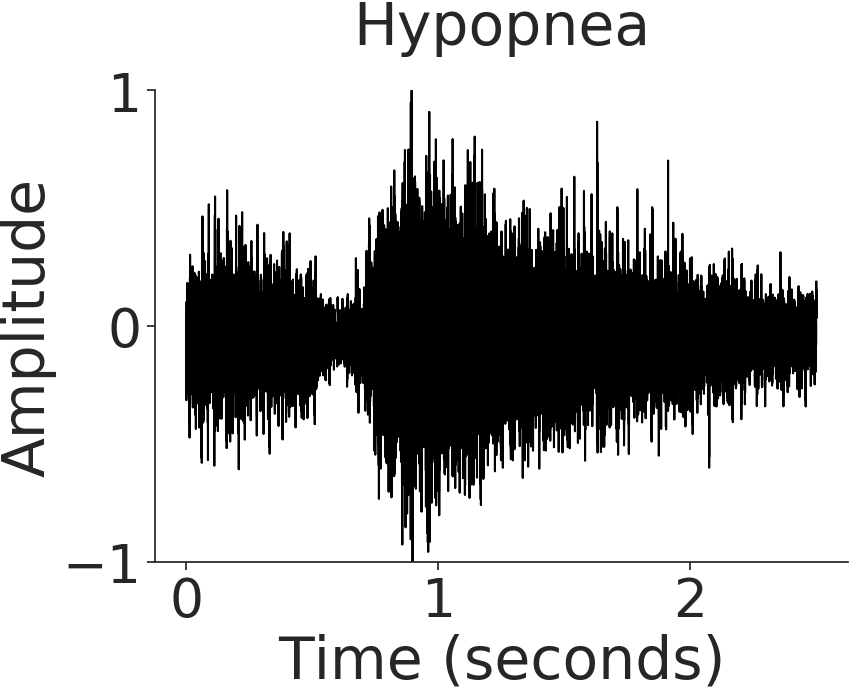}
    \includegraphics[width=.24\textwidth]{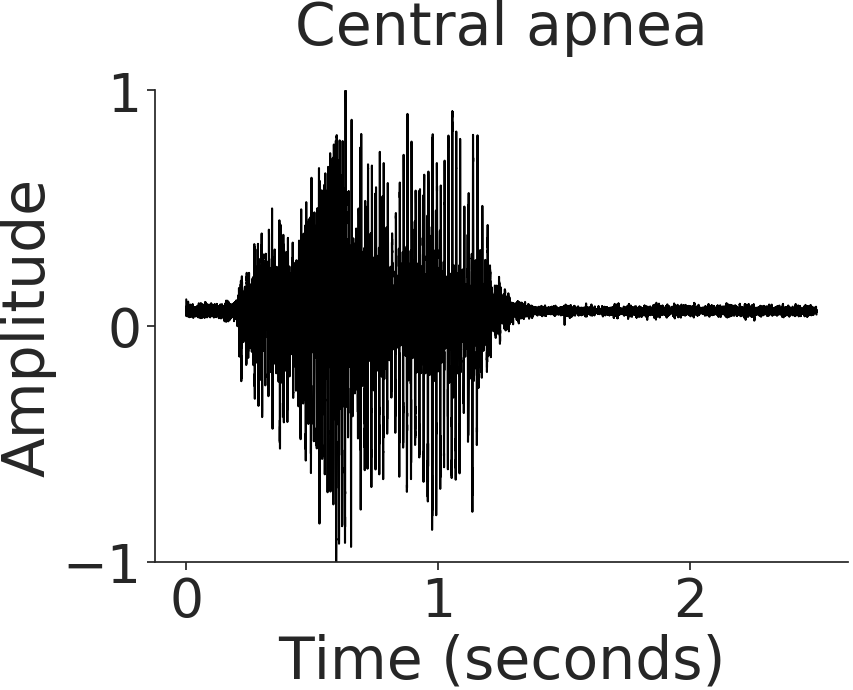}
    \includegraphics[width=.24\textwidth]{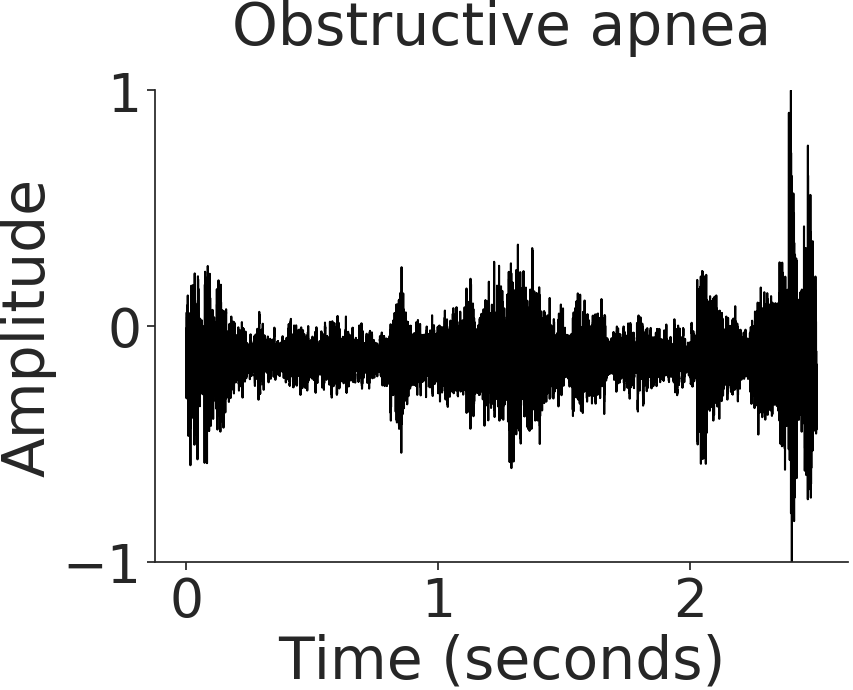}
    \includegraphics[width=.24\textwidth]{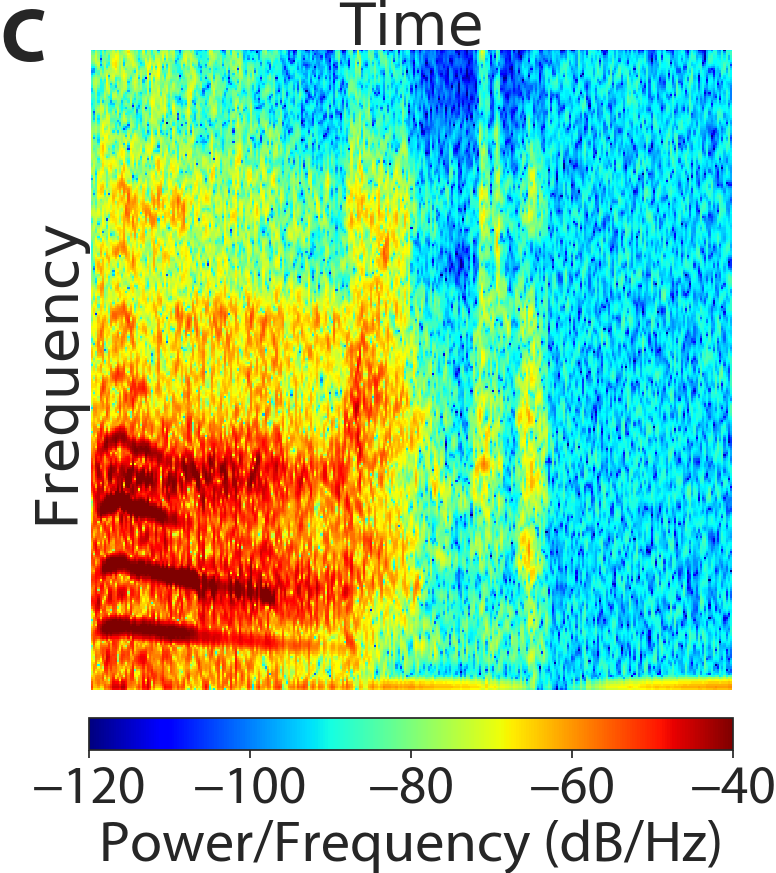}
    \includegraphics[width=.24\textwidth]{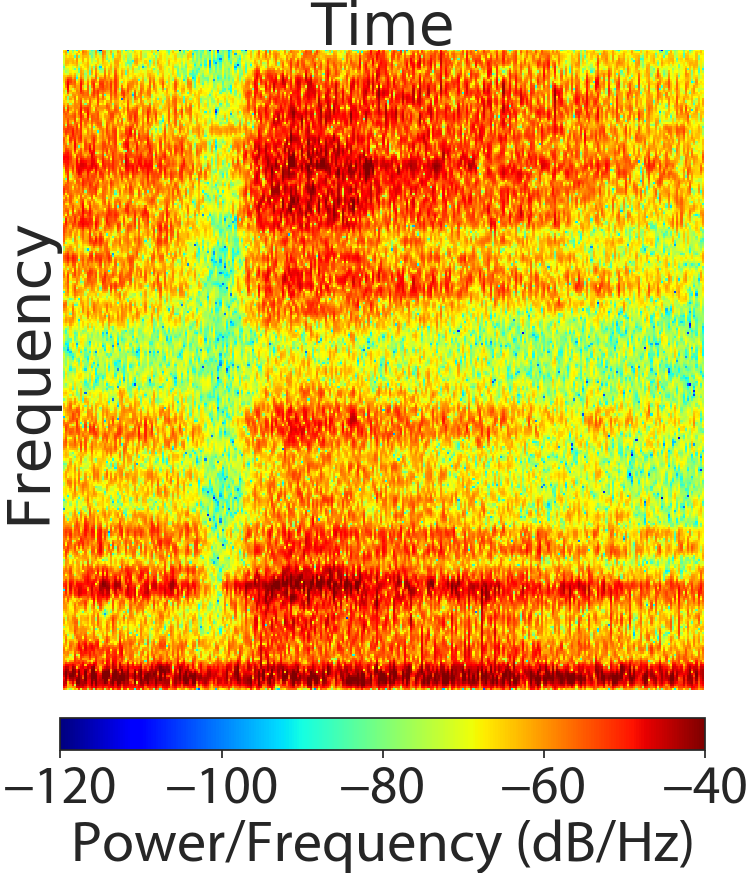}
    \includegraphics[width=.24\textwidth]{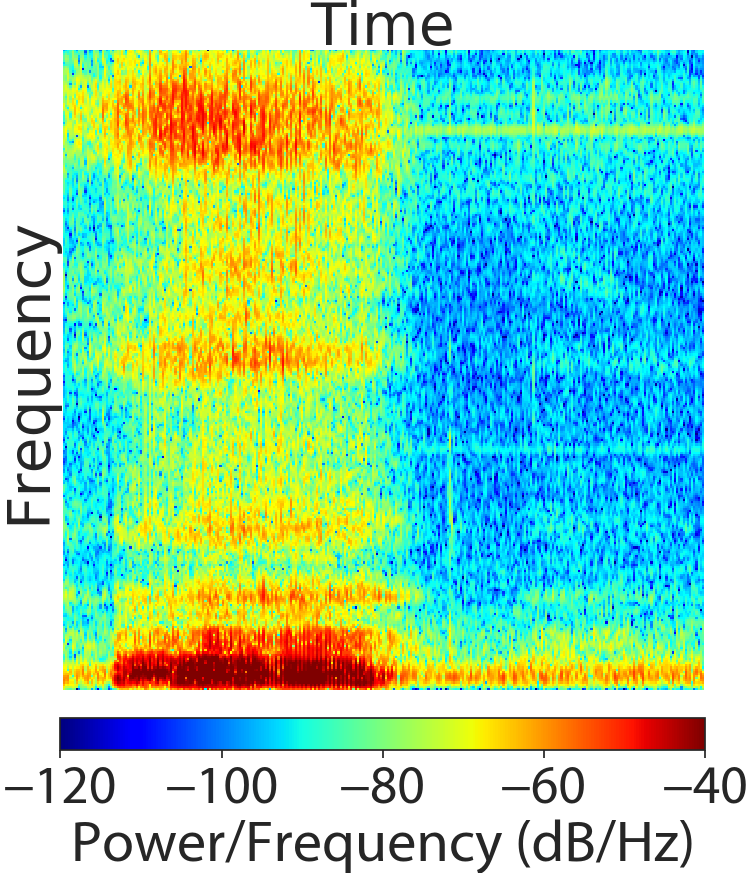}
    \includegraphics[width=.24\textwidth]{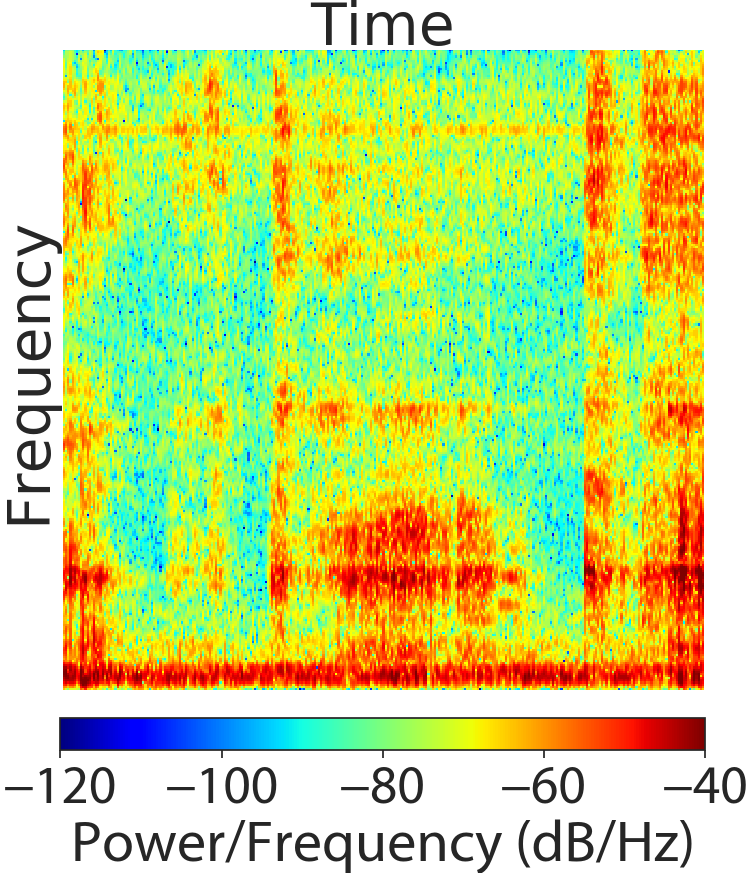}
    \caption{{\bf (A)} Agonal breathing detection pipeline. {\bf (B)} Audio waveform and {\bf (C)} spectrogram of agonal breathing, hypopnea, central apnea, and obstructive apnea.}
    \label{fig:pipeline}
\end{figure}

\begin{figure}
    \centering
    \includegraphics[width=.49\textwidth]{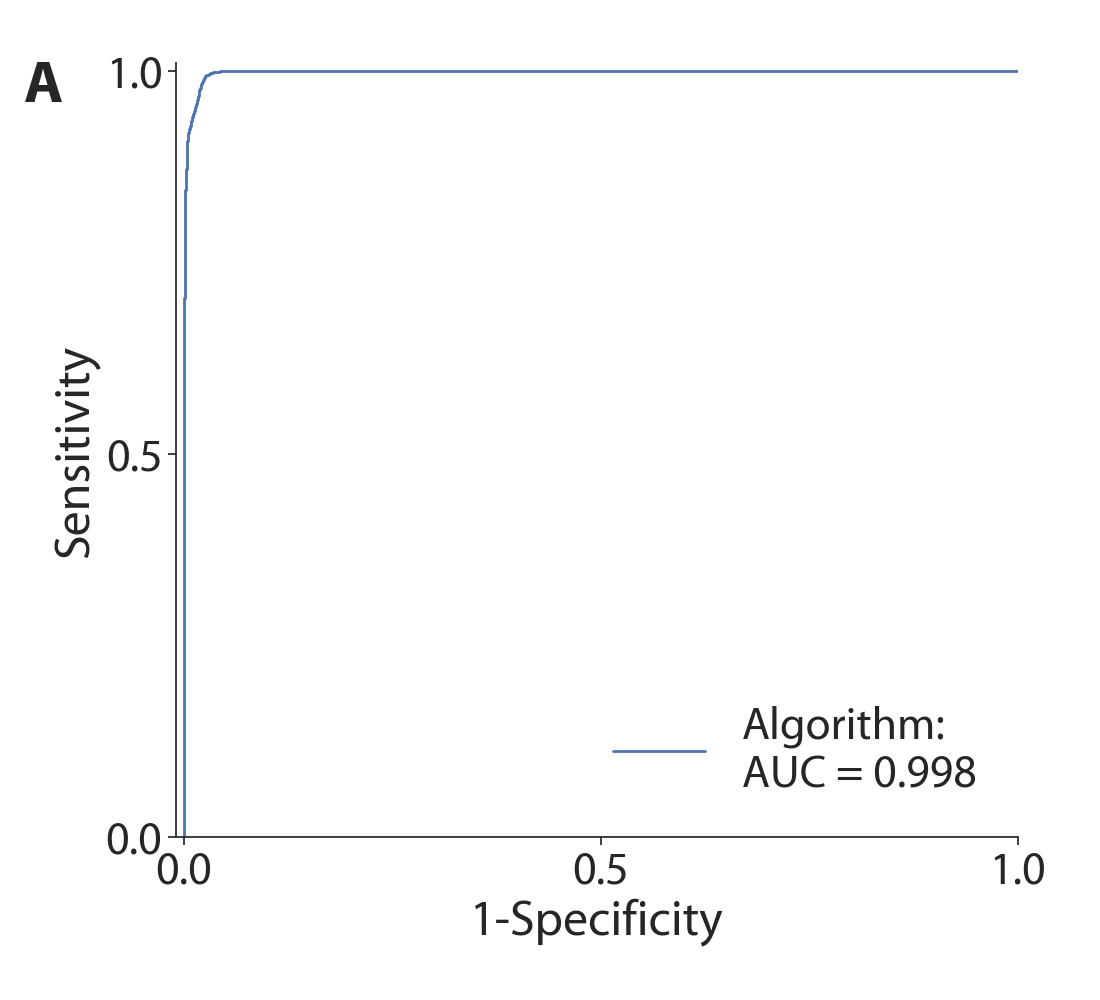}
    \includegraphics[width=.49\textwidth]{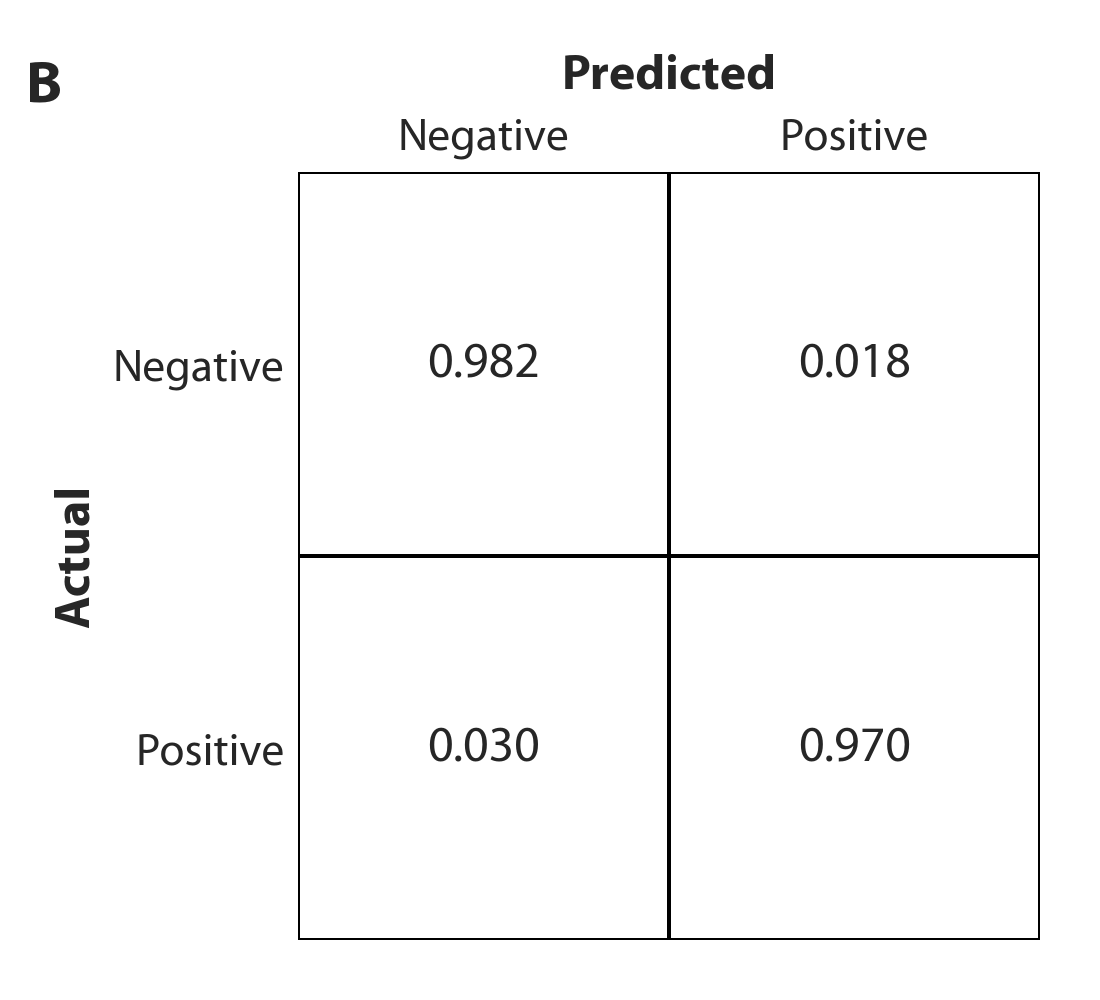}
    \includegraphics[width=.49\textwidth]{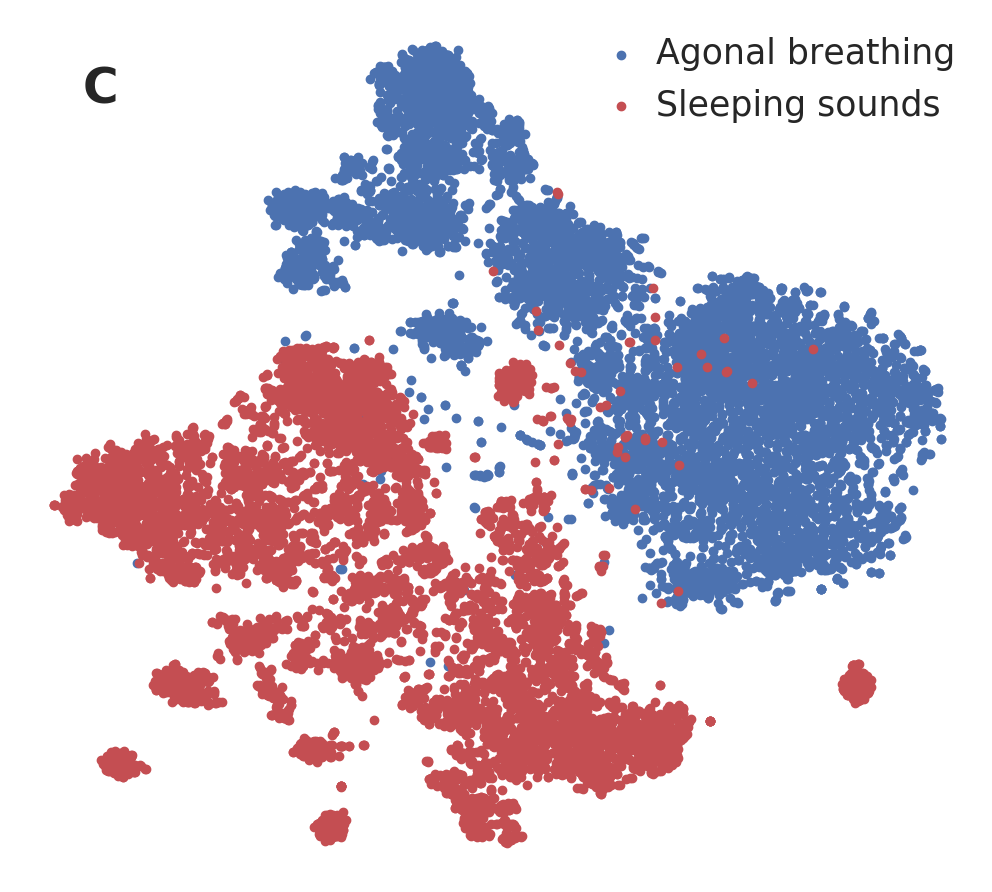}
    \includegraphics[width=.49\textwidth]{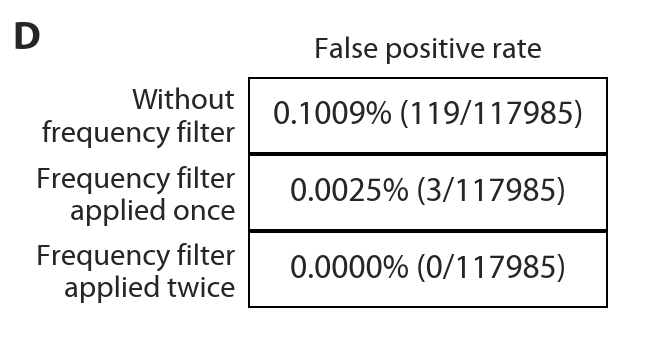}
    \caption{{\bf (A)} ROC curve for our support vector machine classifier, cross-validated on sounds collected during a sleep study and domestic interfering sounds. {\bf (B)} Confusion matrix of agonal breathing and sleeping/domestic interfering sounds indicating the operating point on the ROC curve. {\bf (C)} t-SNE algorithm is applied to visualize the audio embeddings in 2-D. The point clouds show clusters representing the abstract features learned to represent both agonal breathing and negative sound instances. {\bf(D)} The false positive rate when running the classifier across a 82 hour stream of sleep data without and with the frequency filter. By applying a frequency filter to check if the rate of positive predictions matches the rate of agonal breathing, we can decrease the false positive rate.}
    \label{fig:result}
\end{figure}
\section*{Results}

Our agonal breathing recordings are sourced from 9-1-1 emergency calls from 2009--2017, provided by Public Health--Seattle \& King County, Division of Emergency Medical Services. The dataset included 162 calls (19 hours) that had clear recordings of agonal breathing. For each occurrence, we extract 2.5 seconds worth of audio from the start of each agonal breath. We extracted a total of 236 clips of agonal breathing instances. {We augment the number of agonal breathing instances with label preserving transformations, a common technique applied to sparse datasets\cite{cui2015data,wong2016understanding}. We augment the data by playing the recordings over the air over distances of 1, 3 and 6 meters, in the presence of interference from indoor and outdoor sounds with different volumes and when a noise cancellation filter is applied. The recordings were captured on different devices, namely an Amazon Alexa, an iPhone 5s and a Samsung Galaxy S4 to get 7316 positive samples.} Our negative dataset consists of 83 hours of audio data captured during polysomnographic sleep studies, across 12 different patients and includes instances of hypopnea, central apnea, obstructive apnea, snoring and breathing. The negative dataset also includes interfering sounds that might be played in a bedroom while a person is asleep, specifically a podcast, sleep soundscape and white noise. {To train our model, we use one hour of audio data from the  sleep study as well as the other interfering sounds. These audio signals were played over the air at different distances and recorded on different devices to get 7339 samples. The remaining 82 hours of sleep data (117,895 audio segments) is then used for validating the performance of our model.} 

Our model includes audio signals captured by a smart speaker or smartphone and outputs the probability of agonal breathing in real-time on each 2.5s audio segment. We use Google's VGGish model\cite{vggish} as a feature extractor to transform the raw audio waveforms into embeddings which are passed into the SVM. Each segment is transformed from the time-domain into a log-mel spectrogram\cite{muda2010voice}, and is further compressed into a feature embedding using principal component analysis. These embeddings are then passed into an SVM with a radial basis function kernel that can distinguish between agonal breathing instances (positive data) and non-agonal instances (negative data) (Figure \ref{fig:pipeline}A).
Given the relatively small size of our agonal breathing dataset, we augmented the number of agonal breathing instances with label preserving transformations, a common technique applied to sparse datasets\cite{cui2015data,wong2016understanding} (see methods and material). Supplementary Figure \ref{fig:pipeline}B,C show example audio waveforms and spectrograms for agonal breathing, snoring and apnea events.


\subsection*{Classifier Performance}
We applied k-fold (k=10) cross-validation and obtained an area under the curve (AUC) of 0.998 $\pm$ 0.004 (Figure \ref{fig:result}A). We obtain an operating point with an overall sensitivity and specificity of 97.03\% (95\% CI: 96.62 -- 97.41\%) and 98.20\% (95\% CI: 97.87 -- 98.49\%) respectively (Figure \ref{fig:result}B). Our detection algorithm can run in real-time on a smartphone natively and can classify each 2.5 second audio segment within 21 ms. With the smart speaker, the algorithm can run within 58 ms. We visualized the audio embeddings of our dataset by using t-SNE\cite{tsne} to project the features into the 2-D space (Figure \ref{fig:result}C). The two clusters represent the abstract features of agonal breathing instances and audio from the polysomnographic studies.

To evaluate false positive rate, we run our classifier over the full audio stream collected in the sleep lab. The sleep audio used to train the model was excluded from evaluation. By relying only on the classifier's probability outputs, we obtain a false positive rate of 0.1009\%, this corresponds to 119 of 117,985 audio segments. To reduce false positives, the classifier's predictions are passed through a frequency filter that checks if the rate of positive predictions is within the typical frequency at which agonal breathing occurs (i.e., within a range of 3 to 6 agonal breaths per minute\cite{roppolo2009dispatcher,lewis2013dispatcher}). This reduced the false positive rate to 0.0025\%, when considering two agonal breaths within a duration of 10-20s. When considering three agonal breaths within the same duration, the false positive rate reduces to 0\% (Figure \ref{fig:result}D). In our proposed use case a static smart speaker or smartphone would be able to operate on the entire duration of agonal breathing which has been estimated to last for approximately 4 minutes, \cite{roppolo2009dispatcher} in the early phase of cardiac arrest.

\subsection*{Benchmark Performance}
\begin{figure}
    \centering
    \includegraphics[width=.49\textwidth]{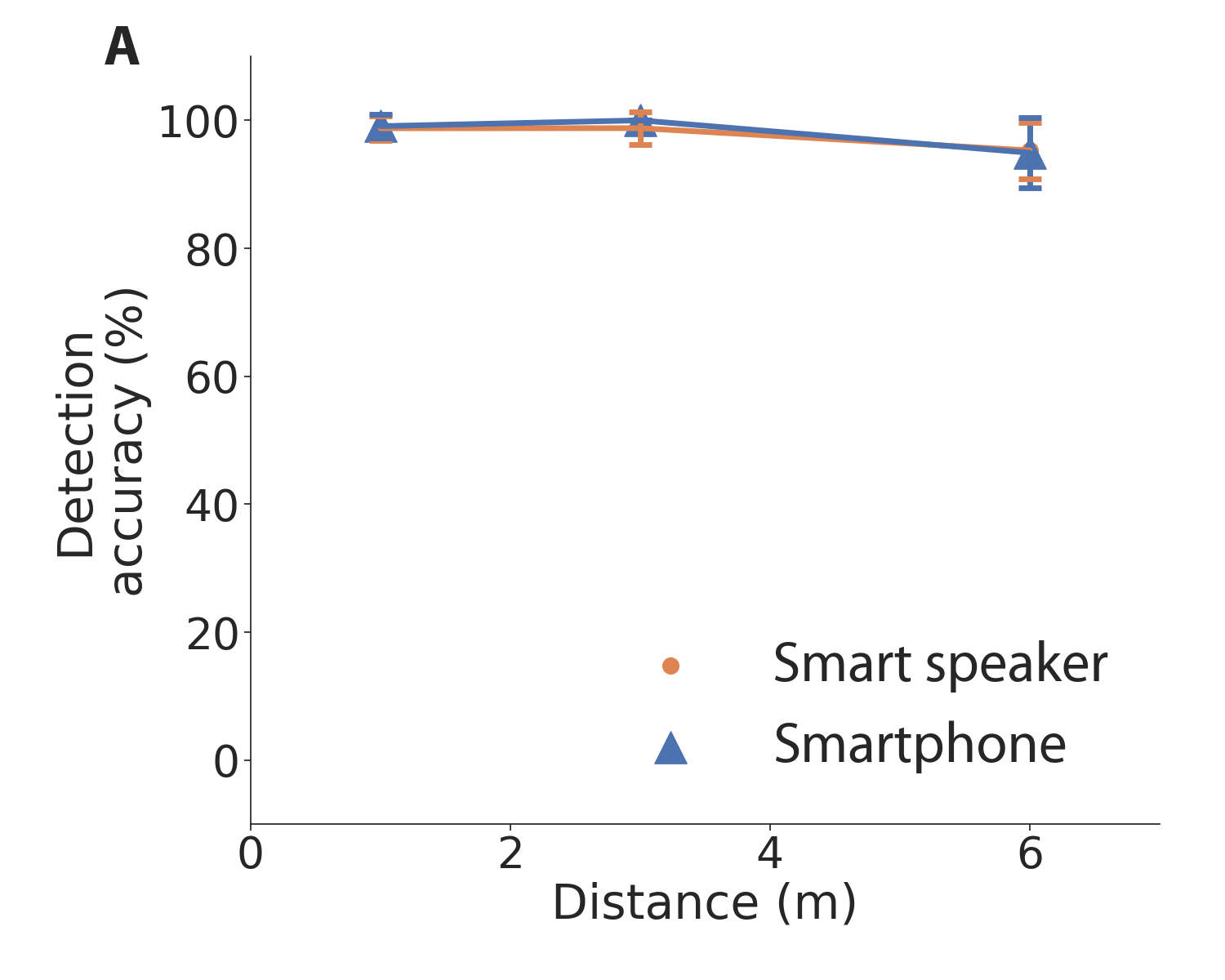}
    \includegraphics[width=.49\textwidth]{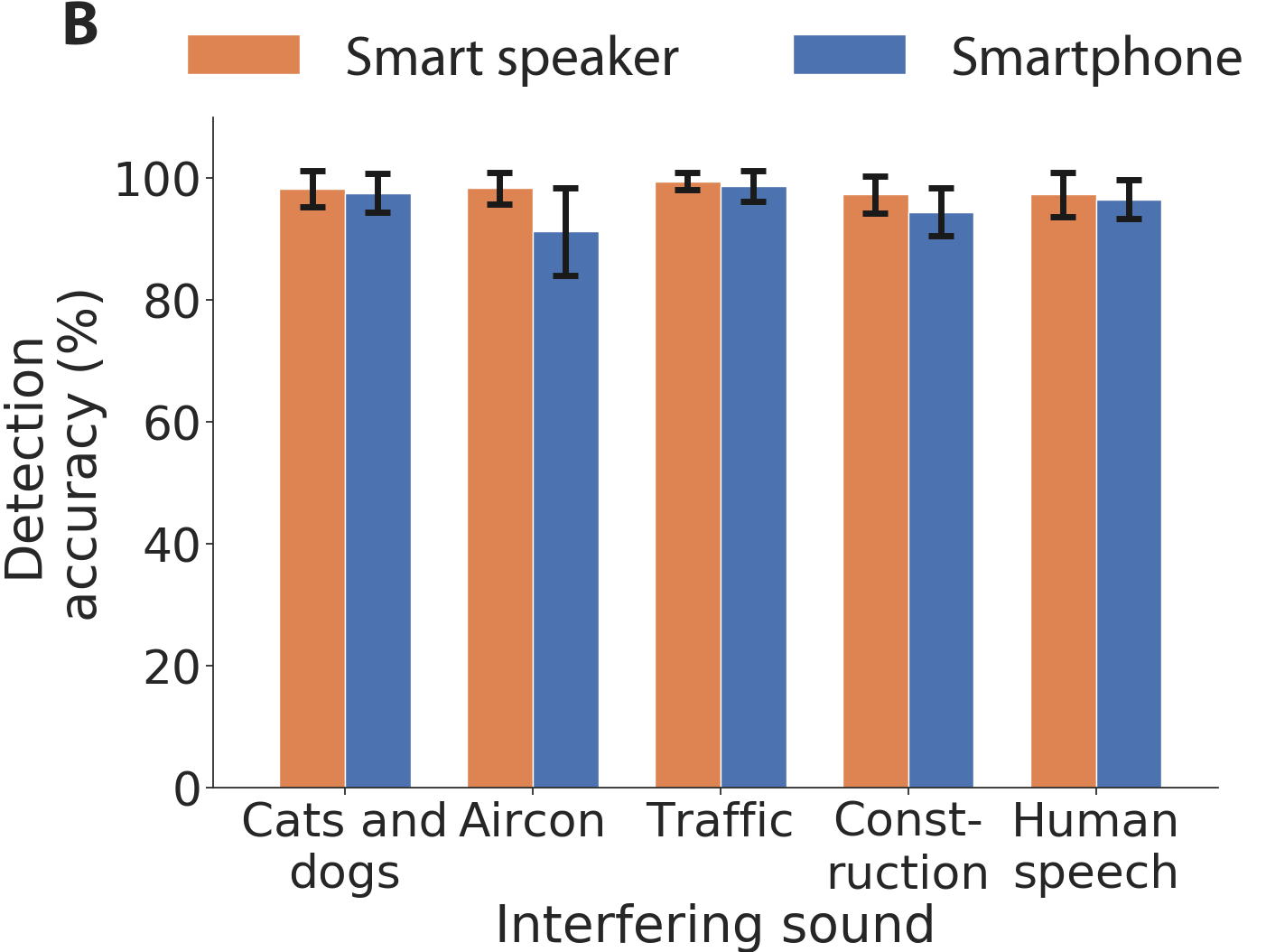}
    \includegraphics[width=.49\textwidth]{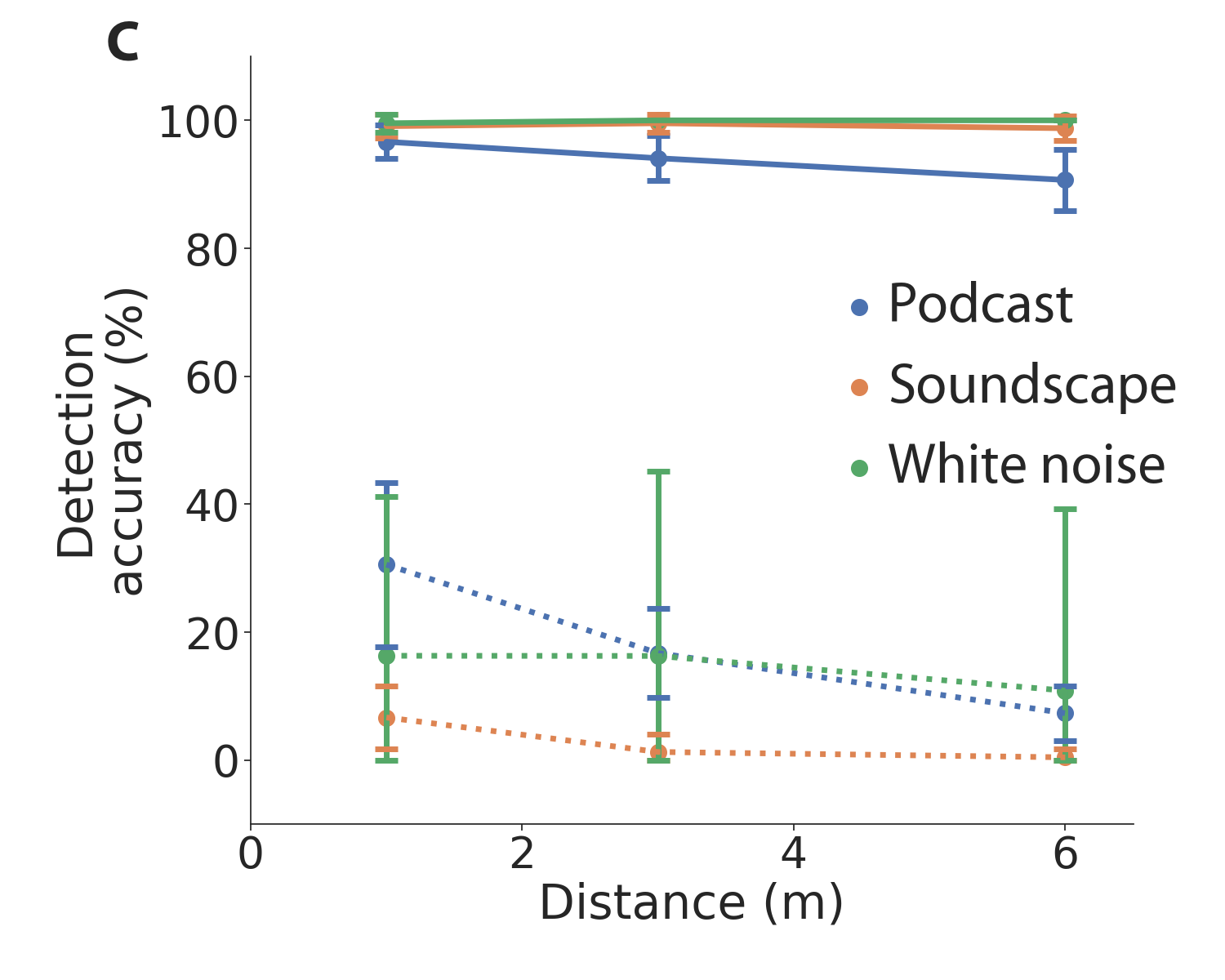}
    \includegraphics[width=.49\textwidth]{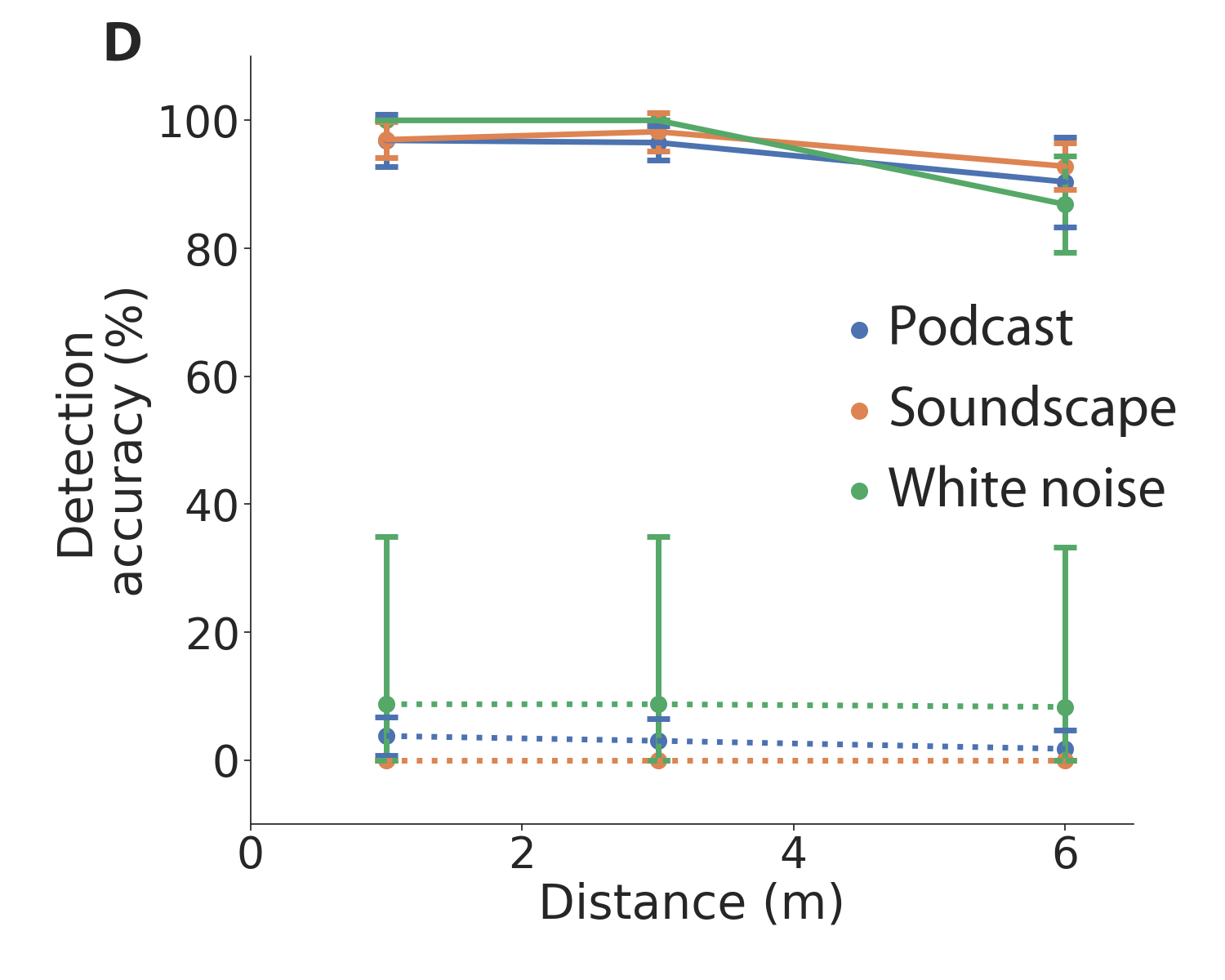}
    \caption{Detection accuracy of smart speaker and smartphone {\bf (A)} across distance {\bf (B)} in the presence of other interfering indoor and outdoor sounds. {\bf (C,D)} With acoustic interference cancellation a smartphone or smart speaker can reduce the effects of its own audio transmissions and become more sensitive at detecting agonal breathing signals. The left and right subplots show the detection accuracy when interfering sounds are played at soft (45 dBA) and loud (67 dBA) volumes respectively. Solid and dashed lines indicate detection accuracy with and without interference cancellation respectively. Error bars indicate the standard deviation accuracy across validation folds.}
    \label{fig:bm}
\end{figure}

We next benchmark the performance of our model.  For these experiments we played the audio clips of agonal breathing over air from an external speaker and captured the audio on an Amazon Echo and Apple iPhone 5s. In Figure \ref{fig:bm} we show the detection accuracy of our classifier in a domestic setting on a smart speaker and smartphone. We evaluate detection accuracy using the k=10 validation folds in our dataset such that no audio file in the validation set appears in any of the different recording conditions in the training set. Figure \ref{fig:bm}A shows the detection accuracy of our classifier in ambient conditions at distances of 1, 3, and 6 meters on the Echo and iPhone 5s. Both devices achieve greater than 98.7\% mean accuracy at distances up to 3 meters. We also evaluated the effect of placing the smartphone in a pocket, with the subject supine on the ground and the speaker next to the head, and obtain a mean detection accuracy of 92.31\% $\pm$ 6.11\%. Figure \ref{fig:bm}B shows our system performance, using the same experimental setup, but in the presence of indoor interfering sounds (cat, dog, air conditioner) and outdoor interfering sounds (traffic, construction and human speech). Across all interfering sound classes the smart speaker and smartphone achieve a mean detection accuracy of 96.95\%. Finally we evaluate how a smartphone or smartspeaker can use acoustic interference cancellation to reduce the interfering effects of its own audio transmissions and improve detection accuracy of agonal breathing (Figure \ref{fig:bm}C,D). We set the smartphone to play sounds one might play to fall asleep including a podcast, sleep soundscape (i.e., river current) and white noise. We play them at a soft (45 dBA) and loud (67 dBA) volume. Simultaneously we play the agonal breathing audio clips. Without any audio cancellation, the detection accuracy is consistently poor, with an average accuracy of 11.84\% and 3.82\% across distances and sounds for soft and loud interfering volumes. When the audio cancellation algorithm is applied, our detection accuracy achieves an average accuracy of 97.60\% and 95.41\% across distances and sounds for soft and loud interfering volumes respectively.

\section*{Discussion}


Out-of-hospital cardiac arrest is a widespread public health concern. Early CPR is a core treatment, underscoring the vital importance of timely detection, followed by initiation of a series of time-dependent coordinated actions which comprise the chain of survival\cite{neumar2015part}. Hundreds of thousands of people worldwide die annually from unwitnessed cardiac arrest, without any chance of survival because they are unable to activate this chain of survival and receive timely resuscitation. Non-contact, passive detection of agonal breathing represents a novel way to identify a portion of previously unreachable victims of cardiac arrest, particularly those who experience such events in a private residence. As the US population ages and more people become at risk for OHCA, leveraging commodity smart hardware for monitoring of these emergent conditions could have public health benefits. The other domains where an efficient agonal breathing classifier could have utility is in unmonitored health facilities (hospital wards and elder care environments\cite{pape2018survival}), EMS dispatch\cite{blomberg2018}, when people have greater than average risk, such as people at risk for opioid overdose-induced cardiac arrest.\cite{Nandakumareaau8914}

An immediate concern of a passive agonal breathing detector is privacy.  For this use case, intentional activation of the device (i.e., ``Hey Alexa" or ``Hey Siri") prior to classification is not feasible because diagnosis involves an unconscious individual in an emergent situation. To address privacy concerns, we envision our system to run locally on the smart devices and not store any data.

An advantage of a contactless detection mechanism is that it does not require a victim to be wearing a device {while asleep in the bedroom, which can be inconvenient or uncomfortable}. Such a solution can be implemented on existing wired smart speakers, and as a result would not face power constraints and could scale efficiently. Potential downsides include that, to date, agonal breathing has been identified in approximately 50\% of cardiac arrest victims, so people experiencing an unwitnessed cardiac arrest without agonal breathing would go undetected by our system. Prior incidence estimates of cardiac arrest-associated agonal breathing events have been based on 9-1-1 calls, which likely biases estimates and underestimates the true incidence of agonal breathing during cardiac arrest.\cite{bobrow2008gasping}

Our study has the following limitations. The number of agonal instances in this study was from one geographic community over an 8-year period. Additional agonal events are needed to assure our model generalizes to variations of agonal breathing. Additional audio of agonal instances, which likely reside in 9-1-1 databases around the world, would also contribute to a more accurate detection system. Also, it is not known if conditions such as seizure, hypoglycemia, severe stroke or drug overdoses with disordered breathing (but not agonal breathing) could be distinct or similar to agonals and thus could pose a challenge to classification. Further work is needed in this area, yet all of these instances represent acute conditions requiring prompt medical intervention. Finally, this proof-of-concept study did not involve EMS activation. A real-world implementation would {sound an alarm and} require a user-interface that provides a cancellation opportunity before the emergency medical response system was activated, so as to further minimize false positives.


Technology is rapidly evolving and in turn providing opportunities to improve human health.\cite{gambhir2018toward,steinhubl2015emerging} The increasing adoption of commodity smart speakers in private residences \cite{gartner} and hospital environments\cite{specialreport} may provide a wide-reaching means to realize the potential of a contactless cardiac arrest detection system.

\section*{Materials and methods}

This study was approved by the University of Washington Institutional Review Board. Data were provided by Public Health--Seattle \& King County, Division of Emergency Medical Services.

\subsection*{Datasets.}
The data represents a subset of 9-1-1 calls which (a) contained a known cardiac arrest and (b) had been identified  to contain cardiac arrest-associated agonal breathing instances. The negative data consists of recordings of 12 patients sleeping in a sleep lab recorded on a Samsung Galaxy S4.

{Our agonal breathing recordings are sourced from 9-1-1 emergency calls from 2009--2017 provided by Public Health--Seattle \& King County, Division of Emergency Medical Services. There are 729 calls totaling to 82 hours. The provided recordings include only calls involving cardiac arrest and specifically those determined to contain occurrences of agonal breathing, either by audible identification of agonal breathing or by description of the breathing from the caller. Each call is further rated by the 9-1-1 operator and a quality assurance reviewer with a confidence score indicating the presence of audible agonal instances. We train our classifier on audio from calls that are rated with high confidence by both the operator and reviewer to contain audible agonal instances. These instances predominantly occur when the 9-1-1 operator asked the caller to place the phone next to the victim's mouth for the purposes of breathing identification. A trained researcher then listened to the 162 calls (19 hours) meeting these criteria and manually recorded timestamps where agonal breathing was  heard during the call (see Supplementary Table 1).  These instances predominantly occur when the 9-1-1 operator asks the caller to place the phone next to the victim's mouth for the purposes of breathing identification. For every timestamp annotation, we extract 2.5 seconds worth of audio from the start of each agonal breath. We extracted a total of 236 clips of agonal breathing instances.}

Our negative dataset consists of 83 hours of audio from polysomnographic studies across 12 different patients and contains instances of hypopnea, central apnea, obstructive apnea, snoring and breathing. The female to male ratio was 0.5 and the median age was 57.0 (IQR: 11.5). The mean number of hypopnea, central apnea and obstructive apnea events across patients was 41, 24 and 26 respectively. The mean apneas-hypopneas index (AHI) is 13, where a value of 0-5 is considered as no apnea, 5-15 is considered as mild apnea and 15-30 is considered as moderate apnea, higher values are considered as severe apnea\cite{nieto2000association}. These annotations were provided by trained sleep technicians. The negative dataset also includes interfering sounds that might be played while a person is asleep: podcast, sleep soundscape and white noise.

We augmented the number of agonal breathing instances with label preserving transformations, a common technique applied to sparse datasets\cite{cui2015data,wong2016understanding}. We augment the data by playing the recordings over the air over distances of 1, 3 and 6 meters, in the presence of interference from indoor and outdoor sounds with different volumes and when a noise cancellation filter is applied. The recordings were captured on different devices, namely an Amazon Alexa, an iPhone 5s and a Samsung Galaxy S4. Similarly, for the negative dataset, portions of the sleep data from all patients were played over the air and recorded on different devices as well as over a phone connection. We play a five minute portion of audio data from each patient over the air at different distances and record the data on an Amazon Alexa, iPhone 5s and over a phone connection.  The entire dataset for cross-validation consists of 14,655 data points with 7316 agonal breathing instances and 7339 instances of negative data. 

\subsection*{Data preparation.} We note that the audio clips were sampled at a frequency of 8~kHz which is standard for audio received over a telephone. All audio clips are normalized between a range of -1 and 1. The audio clips are then passed through Google's VGGish\cite{vggish} model for extracting feature embeddings from an audio waveform. The VGGish model transforms the waveforms into a compact embedding. The model resamples all audio waveforms at 16 kHz then computes a spectrogram using the Short-Time Fourier Transform. A log mel spectrogram is generated and PCA is applied on the spectrogram to produce a 256-dimensional embedding.

\subsection*{Training algorithm.}  We performed k-fold validation (k=10). For any given fold, none of the breathing instances in the validation set occurred in the training set. We evaluate detection accuracy such that no audio file in the validation set appears in any of the different recording conditions in the training set (e.g. if a file played at 6 m is present in the validation set, the same file played at 1 m is not present in the training set). We use a support vector machine with a radial basis function kernel and a regularization (C parameter) of 10. To reduce bias in our classifier we partitioned the data such that recordings from the same call did not straddle the training and validation set split. 

\subsection*{Benchmark experiments.}  To record audio indefinitely on the Echo we used Echo's Drop In feature which streams audio to another smartphone. That smartphone was plugged into a laptop which recorded audio data that was received on the smartphone's audio interface. Audio from the Echo is streamed at 16 kHz and recorded at 44.1 kHz. The iPhone recorded data at 44.1 kHz. Each of the 236 audio clips is prepended with a frequency modulated continuous wave (FMCW) chirp. An FMCW chirp has good auto-correlation properties, as a result we can cross-correlate the recordings from the Echo and iPhone with the chirp to determine the exact time stamp of each audio clip. Each audio clip can then be extracted and transformed into an input for the classifier.

In our benchmark scenarios we evaluate the detection accuracy of our classifier across different distances on a second generation Amazon Echo and an iPhone 5s. We played the 236 audio clips of agonal breathing from a AmazonBasics Wireless Bluetooth speaker and recorded the audio on the Echo and iPhone. The sound intensity of the recordings were approximately 70 dBA at a distance of 1 meter. We fixed the location of the Echo and iPhone and placed the speaker at different distances.

To evaluate the audio interference cancellation algorithm we set the iPhone 5s to play music at two different volumes (45 and 67 dBA), while simultaneously recording audio. We then ran an acoustic interference cancellation algorithm that allowed the smartphone to locally reduce the interference of its own audio transmissions. We used an adaptive least mean squares filter to reduce the dissimilarity between the device's transmission and the received audio recording. Our filter uses the Sign-Data LMS algorithm with 100 weights and a step size of 0.05.

When evaluating system performance in the presence of interfering sounds we use two external speakers, one which plays the agonal breathing recordings and another that plays the interfering noise. The interfering noise is played with a sound intensity of approximately 55 dBA at a distance of 1 meter. 
The interfering sounds are played outside the room containing the agonal breathing speaker and the recording device to simulate sounds that would be heard from outside a bedroom.

\subsection*{Run-time Analysis.}  The most time consuming operations within the detection pipeline are the Fast Fourier Transforms (FFTs) required to generate the spectrogram and running inferences on the audio embeddings. Our iPhone 7 implementation of the detection algorithm used the Accelerate frameworks to perform the FFTs and Monte Carlo sampling to approximate the radial basis function kernel. On an iPhone 7 performing the FFTs to generate a single log mel spectrogram takes 16 ms and running inferences on the support vector machine takes 5 ms. While the classifier can in principle run locally on the Echo device, Amazon currently does not allow third party programs to locally analyse data. Thus, to estimate the performance of our system when run natively on an Amazon Echo, we ran our pipeline on an iPhone 4, which shares the same Cortex-A8 processor as the Echo. On an iPhone 4, computing the spectrogram takes 40 ms and making predictions takes 18 ms.

\section*{Supplementary materials}
Table S1. Demographic summary of sleep study patients with number of hypopnea andapnea events as well as the apneas-hypopneas index (AHI)\\
Table S2. Demographic summary of people with agonal breathing during 9-1-1 calls

\vskip 0.2in
\bibliography{ms}
\bibliographystyle{plain}

\noindent {\bf Acknowledgements.} The authors thank Kevin Jamieson, PhD, Emily Fox, PhD, Gabe Erion, Joseph Janizek, Anran Wang and Samuel Ainsworth for feedback on the manuscript. We thank Rajalakshmi Nandakumar for facilitating the sleep dataset analysis. The authors also thank Sandy Kaplan for editing and Karen Adams and Erik Freidrichsen for facilitating access to the training data. 

\noindent {\bf Author contributions.} SG, JES and JC designed the experiments; JC conducted the experiments; JC and SG designed the algorithms and models; JC conducted the analysis with technical supervision by SG; JC, SG and JES interpreted the results; JES, SG and JC wrote the manuscript; TR provided substantial clinical feedback on the manuscript. JES conceptualized the study using the 9-1-1 dataset and SG conceptualized the sleep dataset analysis.

\noindent {\bf Competing Interests.} All co-authors are inventors on a US provisional patent, submitted by the University of Washington, which is related to this work. JC and SG have equity stakes in Edus Health, Inc., which is not related to the technology presented in this manuscript. SG is a co-founder of Jeeva Wireless, Inc. and Sound Life Sciences, Inc. JES is a co-founder of Sound Life Sciences, Inc.

\noindent {\bf Correspondence.} jesun@uw.edu, gshyam@uw.edu

\setcounter{figure}{0}
\resetlinenumber
\renewcommand{\figurename}{Supplementary Figure}
\renewcommand{\tablename}{Supplementary Table}

\clearpage
\newpage

\section*{\large Supplementary materials}

\vfill
\begin{center}  
  \nolinenumbers \includegraphics [width=\textwidth] {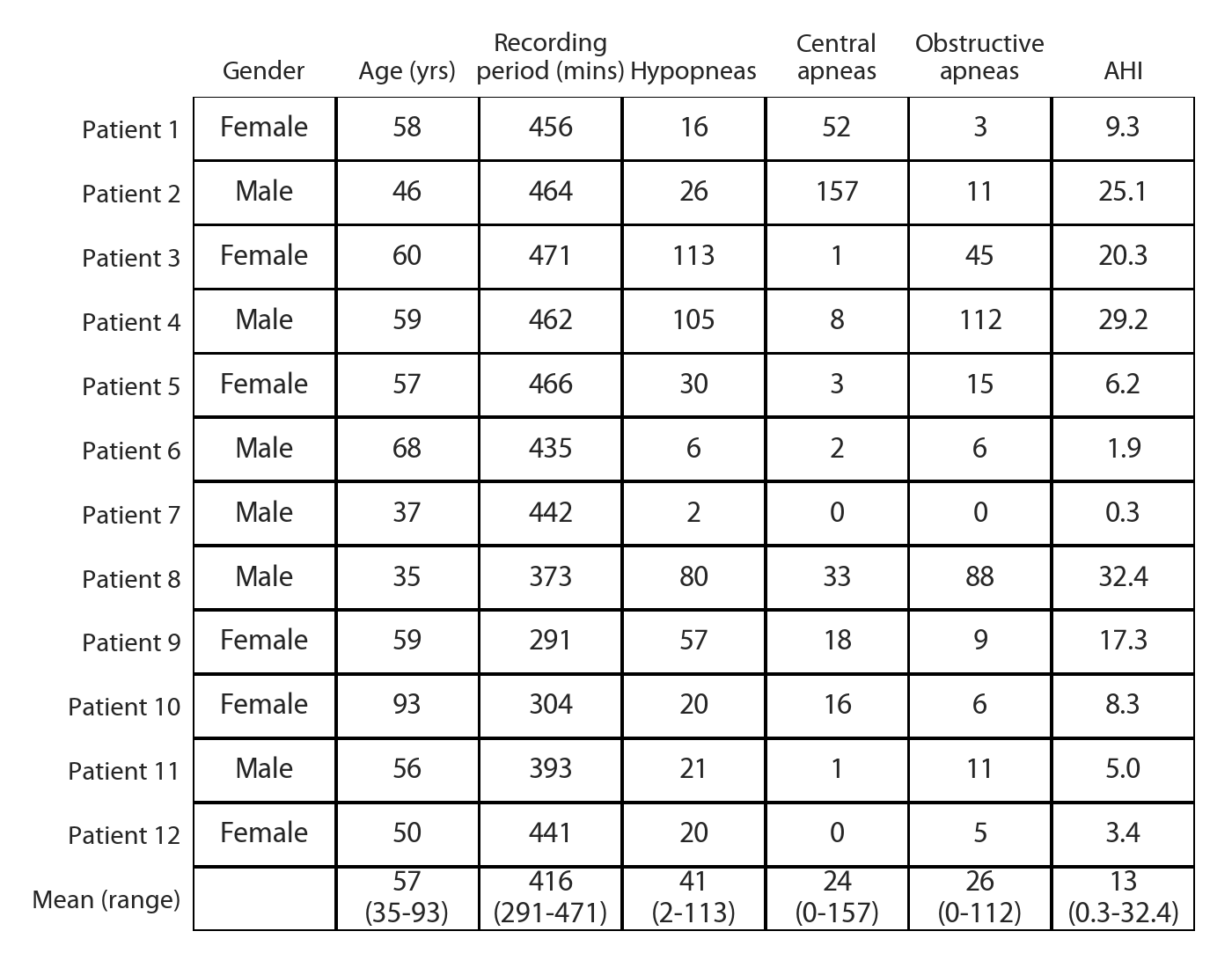}
  \label{fig:sleepdemo}
\end{center}
\nolinenumbers Supplementary Table 1: Demographic summary of sleep study patients with number of hypopnea and apnea events as well as the apneas-hypopneas index (AHI).
\vfill

\vfill
\begin{center}  
  \nolinenumbers \includegraphics [width=.25\textwidth] {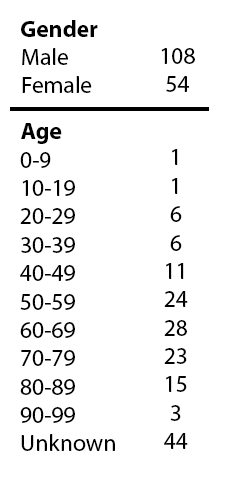}
  \label{fig:sleepdemo}
\end{center}
\nolinenumbers Supplementary Table 2: Demographic summary of people with agonal breathing during 9-1-1 calls.
\vfill

\end{document}